\documentclass[aps,prb,superscriptaddress, showkeys, twocolumn]{revtex4-1}
\usepackage{amsmath,amssymb}
\usepackage{graphicx}
\usepackage{dcolumn}
\usepackage{bm}
\usepackage[latin1]{inputenc}
\usepackage{float}
\usepackage{color}
\usepackage{amsmath}
\definecolor{rojo}{rgb}{1,0,0}
\definecolor{verde}{rgb}{0,0.8,0.5}
\definecolor{azul}{rgb}{0,0,1}
\definecolor{rosa}{cmyk}{0,1,0,0}
\usepackage[toc,page]{appendix} 
\usepackage{latexsym}
\usepackage{amsmath}
\usepackage{mathrsfs}
\usepackage{bm}
\usepackage{amsthm}
\usepackage{physics}
\usepackage{adjustbox}
\usepackage{bbm}
\usepackage{amsfonts}
\usepackage{amssymb}
\usepackage{color}
\usepackage{epsfig}
\usepackage{hyperref}
\hypersetup{colorlinks=true, citecolor=blue}
\usepackage{soul}
\usepackage[normalem]{ulem}
\usepackage[cmtip,all]{xy} 
\usepackage{comment}
\usepackage{multirow}
\usepackage{mathtools}
\newcolumntype{L}{>{$}l<{$}}

\newcommand{\longsquiggly}{\xymatrix{{}\ar@{~>}[r]&{}}} 

\begin{document}

\title{Spin-orbit interaction and spin selectivity for tunneling electron transfer in DNA}

\author{Solmar Varela}
\email{svarela@yachaytech.edu.ec}
\affiliation{Yachay Tech University, School of Chemical Sciences \& Engineering, 100119-Urcuqu\'i, Ecuador}

\author{Iskra Zambrano}
\affiliation{Yachay Tech University, School of Physical Sciences \& Nanotechnology, 100119-Urcuqu\'i, Ecuador}
\author{Bertrand Berche}
\affiliation{Laboratoire de Physique et Chimie Th\'eoriques, UMR Universit\'e de Lorraine-CNRS 7019 54506 Vand\oe uvre les Nancy, France}

\author{Vladimiro Mujica }
\affiliation{School of Molecular Sciences, Arizona State University, Tempe, Arizona 85287-1604, USA}

\author{Ernesto Medina}
\email{emedina@yachaytech.edu.ec}
\affiliation{Yachay Tech University, School of Physical Sciences \& Nanotechnology, 100119-Urcuqu\'i, Ecuador}
\affiliation{Simon A. Levin Mathematical, Computational and Modeling Sciences Center, Arizona State University, Tempe, Arizona 85287-1604, USA}%

\date{\today}

\begin{abstract}
Electron transfer (ET) in biological molecules such as peptides and proteins consists of electrons moving between well defined localized states (donors to acceptors) through a tunneling process. Here we present an analytical model for ET by tunneling in DNA, in the presence of Spin-Orbit (SO) interaction, to produce a strong spin asymmetry with the intrinsic atomic SO strength in meV range. We obtain a Hamiltonian consistent with charge transport through $\pi$ orbitals on the DNA bases and derive the behavior of ET as a function of the injection state momentum, the spin-orbit coupling and barrier length and strength. A highly consistent scenario arises where two concomitant mechanisms for spin selection arises; spin interference and differential spin amplitude decay. High spin filtering can take place at the cost of reduced amplitude transmission assuming realistic values for the SO coupling. The spin filtering scenario is completed by addressing the spin dependent torque under the barrier, with a consistent conserved definition for the spin current.

\end{abstract}

\maketitle

\section{Introduction}
Extensive studies show that electronic transfer in biological systems (for example, photosynthesis and respiration \cite{Winkler1999}) is fast and efficient, which can be explained by means of tunneling processes through organic molecules\cite{Winkler1999,Blumberger}. Hopfield \cite {Hopfield} was one of the first who developed a theory in terms of electron tunneling through a square potential barrier to analyze the electronic transfer in biological systems, finding the expected exponential decrease with the spatial separation of localized states and providing a mechanism to understand the function of the structural characteristics on electron transport molecules. Beratan et al. \cite{Beratan1987, Beratan1991} obtained similar results by showing that the transfer of long-distance electrons in proteins decreases with distance, reinforcing the electron tunneling process as a transport mechanism in these systems. In addition, they showed that the dependence on distance in proteins is related to the structure and that tunneling is mediated by consecutive electronic interactions between connecting donor with acceptor sites.
Electron transfer by pure quantum tunneling have been shown to occur over distances between 20-40 \AA\cite{Gray,Murphy} in biological molecules such as proteins and DNA or $\pi$ conjugated structures. Such processes are temperature insensitive indicating that they are not activated and are partially coherent\cite{TaoMujica}.

Spin active tunneling in chiral molecules has not received considerable attention in spite of its relevance. Recently, Michaeli and Naaman\cite{Michaeli} considered tunneling through the dipole potential produced by hydrogen bonding in a helical geometry. This is a very important model since it is akin to both DNA, oligopeptides with an $\alpha$-helix, strong spin polarizers due to the {\color{black}Chiral-Induced Spin Selectivity (CISS)} effect\cite{Carmeli2,Carmeli1,Xie,Mishra}. Nevertheless, their model does not discuss the details of tunneling coupled to the SO interaction so each component propagates equally through the dipole barrier.

Tunneling processes coupled to spin-activity has been modelled previously for the case of time reversal symmetry breaking. Buttiker\cite{Buttiker} proposed a model for a spin active barrier that considers a magnetic field under the barrier to study the polarization of the transmitted waves and the characteristic dwell times for each spin component. Spin polarization is obtained because the decay strength is spin dependent due to the Zeeman energy contrast in the magnetic field. This mechanism is quite artificial for real molecules and it is in any case very weak but it suggests a similar mechanism using the SO splitting energy basis of the CISS effect. Here, we propose to extend the Buttiker model to the spin-orbit Hamiltonian previously concocted for DNA\cite{Varela2016}, including realistic assumption of a small doping of either electrons or holes by the surface-molecule contact. We find that, analogously to the mechanism found by Buttiker, the energy splitting associate with the spin-orbit term generates different decay rates for each spin species. The different rates produce an exponentially large polarization effect albeit the proven value of the coupling in the meV range\cite{VarelaHydrogenBond}.





This work is organized as follows: In section~\ref{sectionII} we depart from the Hamiltonian model of reference \cite{Varela2016} describing $\pi-\pi$ coupling between neighboring DNA bases including intrinsic SO pathways. A small doping is assumed to tune linear and quadratic terms in the Hamiltonian in reciprocal space. In sections \ref{sectionIII} and \ref{sectionIV} we solve for the tunneling problem with the derived Hamiltonian and discuss the spin-polarization as a function of the tunneling length and SO coupling strength. We also discuss how the torque term in the spin continuity equation that accounts for spin polarization for a time reversal symmetric potential. We end with the Summary and Conclusions.

\section{Molecular Hamiltonian}\label{sectionII}
The full model Hamiltonian for DNA, incorporating the Stark effect for electric fields along the axis of the molecule and atomic Spin-orbit coupling, has been derived recently by Varela et al\cite{Varela2016}. The model involves the orbital basis \{$p_x,p_y,p_z,s$\} on each base on a single helix, assuming weak coupling to the partner strand. The Fermi level for one orbital per base would be at half filling, while light doping of the molecule by electrons or holes e.g. from contact with a substrate, determines the dispersion relation around the Fermi energy.{\color{black} We clarify that mobile electrons in the bases come from $\pi$ orbitals\cite{Simserides}. While these orbitals maybe thought of as fully filled, interactions with neighboring bases and surrounding environment will transfer electrons, process that we model as a change in the filling of these orbitals. The same model will result, perturbatively if we assume half filling and dope with electrons or if we assume fully filled orbitals and dope with holes.}

\begin{figure}[h]
\begin{center}
\includegraphics[width=8cm]{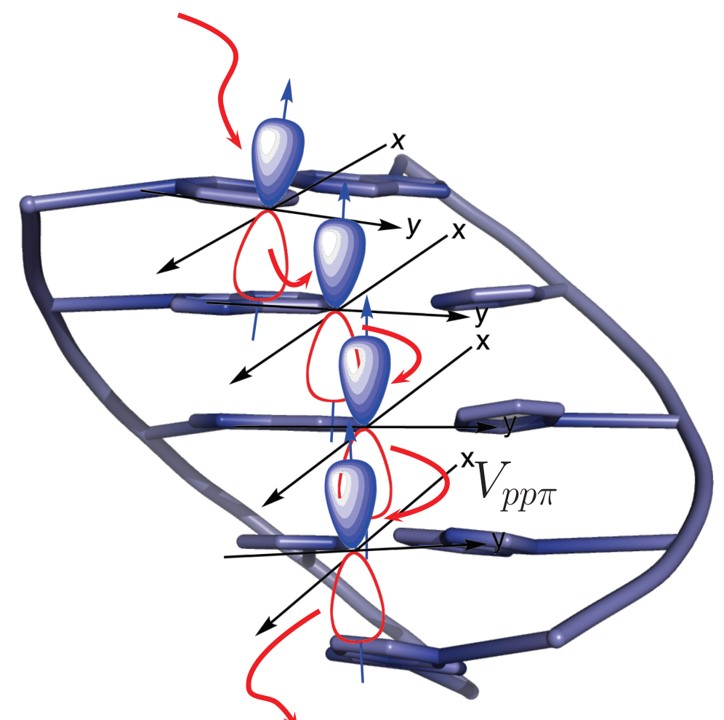}
\caption{Orbital model for transport in DNA. The figure depicts the electron carrying orbitals ($p_z$ orbital perpendicular to the base planes) coupled by $V_{pp\pi}$ Slater-Koster matrix elements. It is well known that any transport mechanism occurs by electron transfer through these orbitals\cite{Simserides}.}
\label{MolecularModel}
\end{center}
\end{figure}

Figure \ref{MolecularModel} shows the $\pi$-stacking model\cite{Genereux} for the double helix, showing only single $p_z$ orbital standing from the basis pairs. The wavefunction overlaps and Spin-Orbit (SO) couplings are derived from a tight-binding Slater-Koster analytical approach with lowest order perturbation theory\cite{Varela2016}. The helical/chiral structure results in a first order SO coupling akin to that of carbon nanotubes\cite{Ando}. The dependence on the chirality and pitch of the helix is built into the SO coupling parameter in the Bloch Hamiltonian.

{\color{black}The largest contributions to the Hamiltonian, considering only the intrinsic spin-orbit coupling of atoms involved in $\pi$ orbitals (N, C, O), comprises two terms
\begin{eqnarray}
H=(\varepsilon_{2p}^{\pi} + 2tf(k)){\bm 1}_s-2g(k)\lambda_{SO}{\bm s}_y,
\label{Hamiltonian1}
\end{eqnarray}
where ${\bm 1}_{s}$ represents the unit matrix in spin space and ${\bm s}_y$ is the Pauli matrix representing the spin degree of freedom in the local coordinate system of the molecule. 
The first term in the Hamiltonian (\ref{Hamiltonian1}), involves the base $2p-\pi$ orbital energy ($\varepsilon_{2p}^{\pi}$) and the kinetic energy with $t$, diagonal in spin space, depending on explicit structural parameters of the molecule
\begin{equation}
t=V_{pp}^\pi+\frac{b^2\Delta\phi^2\left (V_{pp}^\sigma-V_{pp}^\pi\right)}{(8\pi^2r^2(1-\cos\Delta\phi)+b^2\Delta\phi^2)},
\end{equation}
where $\Delta{\phi}$ is the twist angle per base, $r$ and $b$ are the radius and pitch of the helix, respectively and $V_{pp}^{\sigma,\pi}$ are the Slater-Koster overlaps between consecutive bases. 

The second term in (\ref{Hamiltonian1}), is the spin active term
\begin{equation}
\lambda_{SO}=\frac{4\pi\xi_p r b\Delta\phi(1-\cos\Delta\phi)\left (V_{pp}^\sigma-V_{pp}^\pi\right)}{(\varepsilon_{2p}^\pi-\varepsilon_{2p}^\sigma)(8\pi^2r^2(1-\cos\Delta\phi)+b^2\Delta\phi^2)},
\end{equation}
where $\xi_p$ is the atomic SO coupling of double bonded atoms in the bases (either C, O or N) and $\varepsilon_{2p}^{\pi,\sigma}$ are the bare energies if the $2p$ valence orbital is either $\pi$ (perpendicular to the base) or $\sigma$ bonded. The $\lambda_{SO}$ parameter, like the $t$, also includes all geometrical characteristic of the helix. Finally, $f(k)=\cos({\bm k}\cdot {\bm R})$ and $g(k)=\sin({\bm k}\cdot {\bm R})$ are the functions of reciprocal space with $R$ the lattice parameter and $k$ the wave vector in the local system of the helix. This Hamiltonian only includes the dominant spin active terms derived from the geometry dependent spin-orbit coupling. Additional spin active terms are three to six orders of magnitude smaller\cite{Varela2016}}. 

The previous simplified model is based on the basis set \{$p_x,p_y,p_z$\} orbitals per DNA base so half filling is assumed ($p_x,p_y$ are $\sigma$-bonded while $p_z$ is single filled)\cite{Varela2016}. Charge transfer/doping by the environment of the molecule or by the substrate on which the molecule is attached, can add or subtract charge shifting the dispersion from the inflection point $K_{\mu}$ for the purely kinetic Hamiltonian. Thus, the Fermi energy corresponds to $K_{\mu}=0$, so that $k=K_{\mu}+q$ describes a perturbative doping in the vicinity of the Fermi level. Expanding to lowest order in $q$ and assuming that $\bm{k}\cdot\bm{R}\ll 1$ we have
\begin{eqnarray}
f(k)&=&1-\frac{q^{2}R^2}{2}+{\cal O}(q^4)...~~,\notag\\
g(k)&=&qR+{\cal O}(q^3)...,
\end{eqnarray}
and the resulting Hamiltonian is: 
\begin{equation}
H=\left[\varepsilon_{2p}^{\pi}+2t\left(1-\frac{q^{2}R^2}{2}\right)\right]{\bm 1}_s - 2qR\lambda_{SO}{\bm s}_y.
\label{HamiltonianK1}
\end{equation}
Note that to lowest order in $q$ we have a quadratic dispersion for the kinetic term and a linear dispersion for the spin-orbit term. 

In the sense of $k\cdot p$ theory\cite{WinklerSpin} we can requantize this Hamiltonian to treat the tunneling problem in the vicinity of the Fermi level: $q\longrightarrow -i\partial_{x}$ and $-\hbar^{2}/{2R^2|t|}\longrightarrow m$.
Eliminating constant energy terms we arrive at
\begin{equation}
H=\frac{1}{2m}(-i\hbar\partial_{x})^{2}\mathbf{1} +\alpha\sigma_{y} ({-i}\hbar{\partial_x}),
\end{equation}
where $\alpha=-\frac{2R}{\hbar}\lambda_{SO}$ and $\sigma_y$ is the Pauli matrix. This derivation results in the same Hamiltonian surmised in reference\cite{MedinaGonzalez} and leads to the detailed physics of the {\color{black} CISS effect} in the absence of tunneling. The Hamiltonian in reference \cite{Varela2016} can then be considered as a microscopic derivation of the continuum description. Note also that in the model recently proposed for Helicene \cite{GutierrezHelicene}.

\section{Potential barrier}\label{sectionIII}
\begin{figure}[h]
\begin{center}
\includegraphics[width=8cm]{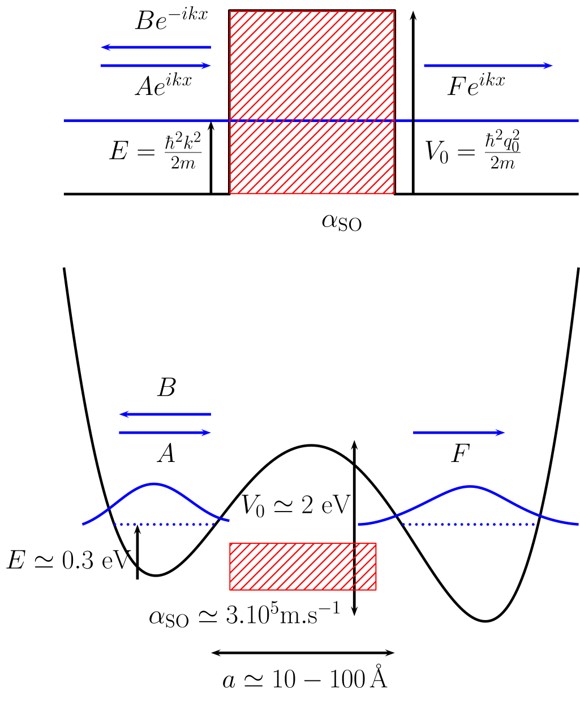}
\caption{{\color{black}Scattering potential barrier model with SO interaction (red hatch). The label for the incident ($A$) and scattered ($B$ and $F$) wave functions amplitudes are indicated.  The well parameters are estimated in the text on the basis of polaron transport.} }
\label{SistemaTunel}
\end{center}
\end{figure}

We now introduce the previous model under a potential barrier assuming, as shown in Figure~\ref{SistemaTunel}, that electrons are injected from (and partially reflected back to)  a donor localized state and received at an acceptor site. One might also consider dipole barriers as expected from hydrogen bond generated potential identified in reference\cite{Michaeli}. We consider an incident state of momentum $p_x$, where $x$ is along the helix tangent. Electrons interact with a potential barrier of height $V_0$ and width $a$. In the barrier region the SO interaction is active (see Fig.\ref{SistemaTunel} and reference~\cite{Cribb}). 

The scattering problem is then defined by
\begin{equation}
H=\left\{
\begin{array}{ccl}
\left(\frac{p_x^{2}}{2m} + V_{o}\right)\mathbf{1} +\alpha\sigma_{y} p_{x} & ; & 0 \leqslant x  \leqslant a\\
{\rm donor}/{\rm accept.~~states}& ; & \hbox{outside}
\end{array}
\right.
\label{HamiltonianFull}
\end{equation}
The parameters used for the injected momentum, barrier height and the range of spin-orbit values are selected as follows: by electron transfer from experimental techniques based on coupling artificial donor and acceptor sites has been tested using a series of well-conjugated molecules including metallo-intercalators, organic intercalators, organic end-cappers\cite{Intercalators}. Measurements, using DNA as a bridge, report tunneling between 10-40 \AA \cite{BeratanTunnel,Gray,Murphy,Barton}. On the other hand the barrier heights reported are in the range of 0.5-2.5 eV\cite{Waldeck,Cuniberti}, either by the potential difference between the metal intercalators as donors/acceptors or the substrate in an STM setup,  and the HOMO state of Guanine\cite{Roche}.   

The donor confinement potential can give an idea of the approximate $k$ vector values being injected into the barrier, assuming carriers are in the ground state. In intercalators such as those in reference\cite{Barton} or STM setups \cite{TaoMujica} report confinement over one or two base pairs. We can estimate $k=0.4$ nm$^{-1}$, {\color{black} corresponding to the incident energy of $E=0.24$ eV, and  $V_0=2$ eV.} With these experimentally derived parameters, we can see the consequences of differential spin tunneling with the derived Hamiltonian.

\section{Tunneling problem}\label{sectionIV}
Once we have estimated the barrier parameters and the polaron well parameters,
we can fully solve the 1D scattering problem by assuming an initial pure spin state. To determine the scattering properties we can solve the problem with simple plane wave injection conditions. The Hamiltonian $H$ acts on spinors $\psi$ with the form
\begin{equation}
\psi(x)=\left(
\begin{array}{c}
\psi_{\uparrow}(x)\\
\psi_{\downarrow}(x)
\end{array}
\right),
\end{equation}
{\color{black}where the arrows indicate the spin components}. If the incident beam is given by 
 \begin{equation}
 \psi_{in}(x)=\left(\begin{array}{c}
 A_{\uparrow}\\
 A_{\downarrow}
 \end{array}
 \right)e^{ikx}, 
 \end{equation}
and the spinor for the scattered beam is
\begin{equation}
\psi_{out}(x)=\left(\begin{array}{c}
 F_{\uparrow}\\
 F_{\downarrow}
 \end{array}
 \right)e^{ikx},
\end{equation}
then, the spin asymmetry of the scattered beam cam be written in the form
\begin{equation}
P_z= \frac{|F_{\uparrow}|^2-|F_{\downarrow}|^2}{|F_{\uparrow}|^2+|F_{\downarrow}|^2}.
\label{Asymmetry}
\end{equation}


Now, Considering an incident electron with energy $E$ and wave vector $k$, the general solutions are
 
\begin{equation}
\begin{array}{ll}
\psi_{\rm 1}=\left(\begin{array}{c}
A_{\uparrow} \\
A_{\downarrow}
\end{array}\right)
e^{ik_1x}+\left(\begin{array}{c}
B_{\uparrow} \\
B_{\downarrow}
\end{array}\right)e^{-ik_1x}; & x\leqslant0
\end{array},
\label{Sol01}
\end{equation}

\begin{equation}
\begin{array}{lcl}
\psi_{\rm 3}=\left(\begin{array}{c}
F_{\uparrow} \\
F_{\downarrow}
\end{array}\right)e^{ik_3x}&;
 & x\geqslant a
\end{array},
\label{Sol03}
\end{equation}
and in the region of the barrier, the solution when $E>V_0$ is 
\begin{equation}
\begin{array}{ll}
\psi_{\rm 2}=\left(\begin{array}{c}
C_{\uparrow}e^{iq_{\uparrow}x} \\
C_{\downarrow}e^{iq_{\downarrow}x}
\end{array}\right)
+\left(\begin{array}{c}
D_{\uparrow}e^{-iq_{\uparrow}x} \\
D_{\downarrow}e^{-iq_{\downarrow}x}
\end{array}\right); & 0\leqslant x\leqslant a
\end{array},
\label{Sol02}
\end{equation}
and, for $E<V_0$
\begin{equation}
\begin{array}{ll}
\psi_{\rm 2}=\left(\begin{array}{c}
C_{\uparrow}e^{q_{\uparrow}x} \\
C_{\downarrow}e^{q_{\downarrow}x}
\end{array}\right)
+\left(\begin{array}{c}
D_{\uparrow}e^{-q_{\uparrow}x} \\
D_{\downarrow}e^{-q_{\downarrow}x}
\end{array}\right); & 0\leqslant x\leqslant a
\end{array},
\label{Sol02}
\end{equation}
{\color{black}where $C$ and $D$ are the amplitudes inside barrier region}. 

Solving the eigenvalue problem $H\psi=E\psi$ for each of the regions, we have that wave vectors for the electron in 
1 and 3 are $k_{\rm 1}=k_{\rm 3}=\sqrt{2mE}/\hbar$ and for region 2, the wave vector $q$ depends on the spin orientation and, if $E>V_0$ is given by
\begin{equation}
q_{s}= \sqrt{k^{2} - q_{0}^{2} +\left(\frac{m\alpha}{\hbar}\right)^{2}} +s\left(\frac{m\alpha}{\hbar}\right),
\label{wavevectorI}
\end{equation} 
and if $E<V_0$, then
\begin{equation}
q_{s}= \sqrt{| q_{0}^{2} - k^2| -\left(\frac{m\alpha}{\hbar}\right)^{2}} -is\left(\frac{m\alpha}{\hbar}\right),
\label{wavevectorII}
\end{equation} 
where $q_0^{2}=2mV_{0}/\hbar^{2}$, $k^{2}=2mE/\hbar^{2}$. $s$ is the label associated with the spin up(down) such that $s=+(-)$. One can see the explicit dependence of $q$ with the spin $s$, $V_0$ and with the  SO magnitude, {\color{black}$\alpha$}. {\color{black}Note that if $E>V_0$ then wave vector $q_s$ in the barrier region is real and the amplitudes will oscillate due to standing wave patterns between the edges of the barrier and the spin precession (relative changes in the spinor amplitudes) due to the SO coupling}.

The coefficients are determined by the requirement of the continuity of the wave function at $x=0$ and $x=a$  following reference\cite{Molenkamp}: $\psi_{1,s}(0)=\psi_{2,s}(0)$, $\psi_{2,s}(a)=\psi_{3,s}(a)$
and $\hat{v}_{x,1}\psi_{1}(0) = \hat{v}_{x,2}\psi_{2}(0),~~
\hat{v}_{x,2}\psi_{2}(a) = \hat{v}_{x,3}\psi_{3}(a)$
where the velocity $\hat{v}_x=\partial H/\partial p_x$ in regions 1 and 3 have the form
\begin{equation}
\hat{v}_{x,1}=\hat{v}_{x,3}=\left(
\begin{array}{cc}
p_x/m & 0 \\
   0  & p_x/m
\end{array}
\right), 
\end{equation}
and in region 2
\begin{equation}
\hat{v}_{x,2}=\left(
\begin{array}{cc}
p_x/m & -i\alpha \\
  i\alpha  & p_x/m
\end{array}
\right).
\end{equation}

\subsection{Energies below the barrier $E<V_0$}
Below the barrier transmission will be the most common physical scenario where we have an interplay between three energies: i) the incoming energy of the electron estimated by the quantum well that precedes the barrier, ii) the barrier height $V_0$ and iii) the SO energy that has been estimated to be in the meV range \cite{VarelaHydrogenBond}. It is useful to consider some possible values of the wavevector inside barrier $q_s$ (Eq.\ref{wavevectorII}):
\begin{itemize}
    \item $\alpha=0$, $q_s=\sqrt{|q_0^2 - k^2|}$ and no spin activity is expected. Simple wave function decay is expected.
    \item  $|q_0^2 - k^2|>(m\alpha/\hbar)^2$, then $q_s$ will be a complex number ($\alpha\neq 0$). Then we have an underdamped decay of the barrier wavefunction.
    \item If $|q_0^2 - k^2|<(m\alpha/\hbar)^2$, then $q_s$ is purely imaginary number and the wave function is a plane wave. 
\end{itemize}
 When the spin-orbit energy $E_{SO}$, approaches $|\hbar^2k^2/2m-V_0|$, a transition is expected between the two previous regimes. 

\begin{figure}[H]
\begin{center}
\includegraphics[width=8.8cm]{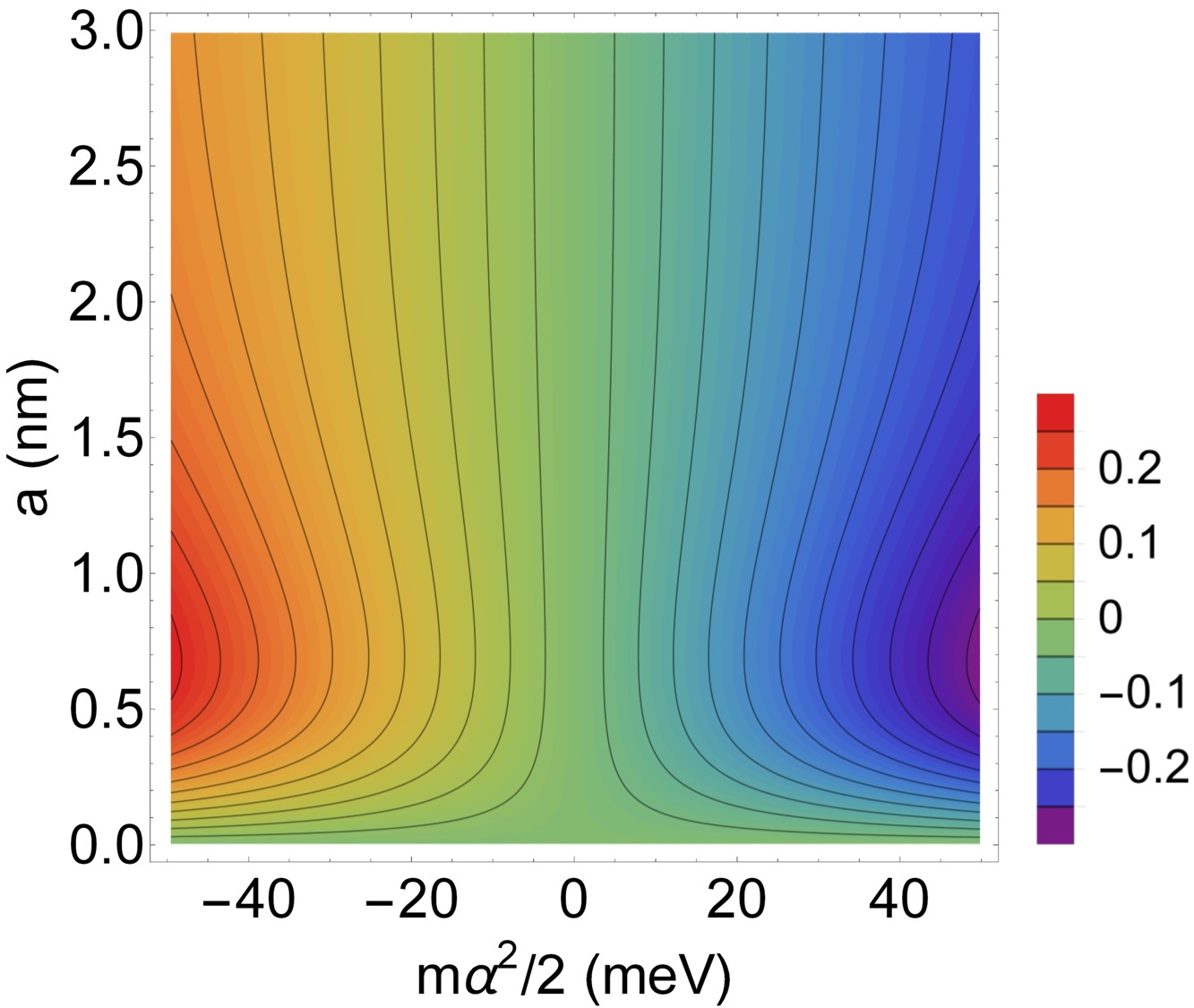}
\caption{\color{black}Spin asymmetry $P_z$ as a function of $a$ in nm and the energy of the SO interaction in meV. The values for the incident wave function of electron $k=0.44$ nm$^{-1}$ and the barrier height $q_0=1.2$ nm$^{-1}$ are fixed.}
\label{BelowAsymmetry1}
\end{center}
\end{figure}

\begin{figure}[H]
\begin{center}
\includegraphics[width=8.8cm]{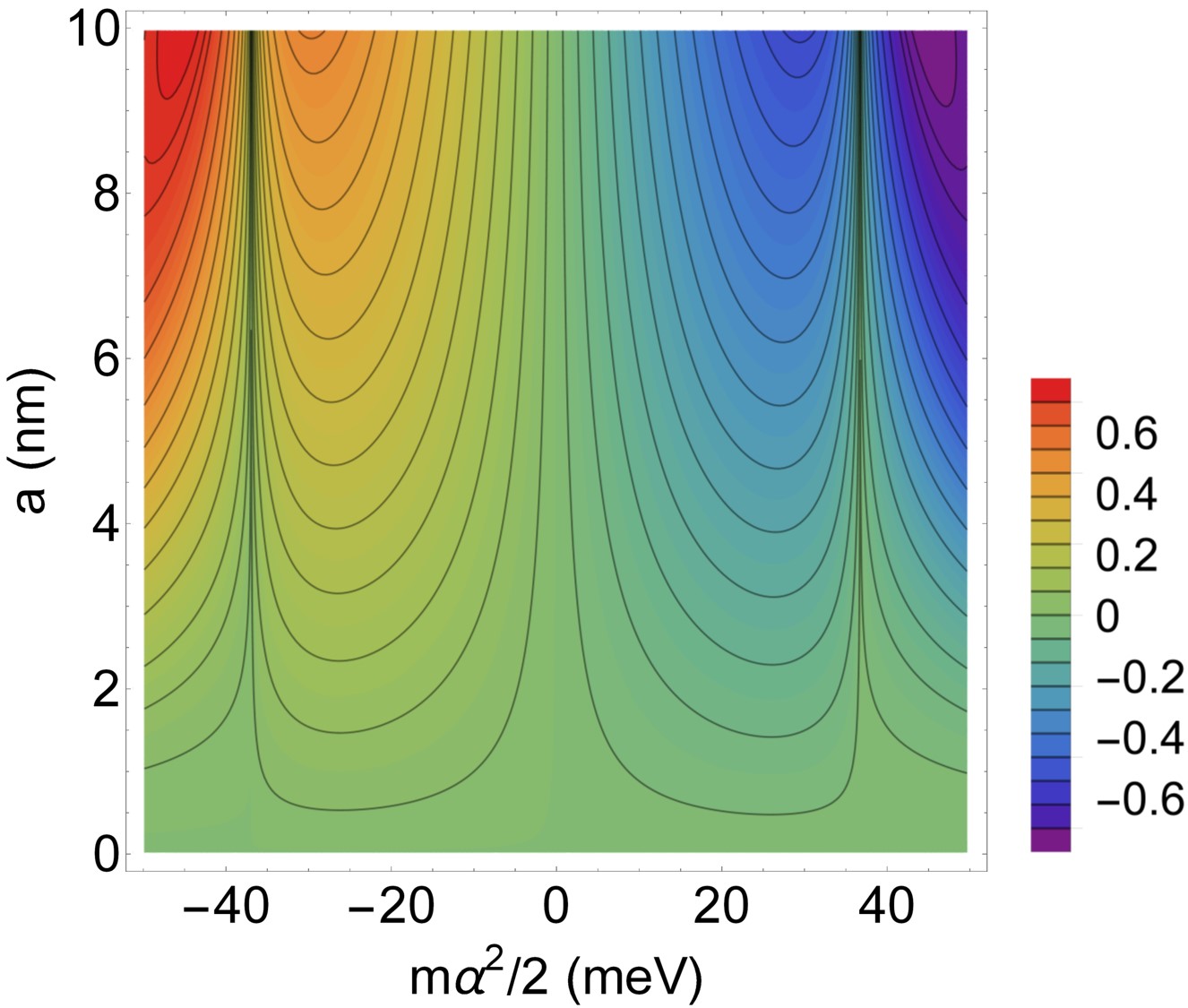}
\caption{Spin asymmetry $P_z$ as a function of $a$ in nm and the energy of the SO interaction in meV. The values for the incident wave function $k=0.44$ nm$^{-1}$ and the barrier height $q_0=0.46$ nm$^{-1}$ are fixed.}
\label{BelowAsymmetry2}
\end{center}
\end{figure}

\begin{figure}[H]
\begin{center}
\includegraphics[width=8.5cm]{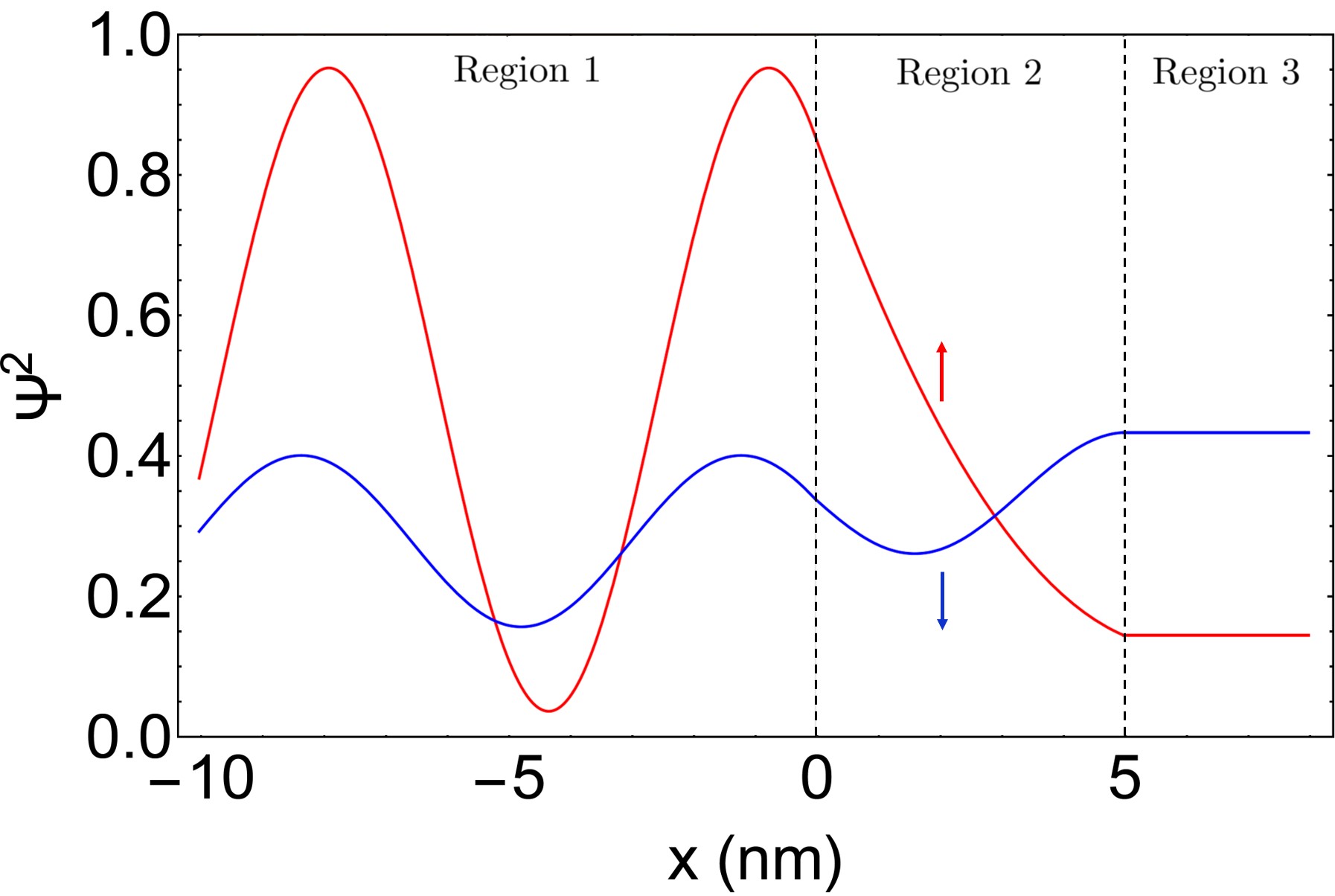}
\caption{Spin component probabilities for energies in the vicinity of the barrier height. The observe oscillation accounts for the reentrant behavior predicted for the polarization. We used $k=0.440$ nm$^{-1}$, $q_0=0.446$ nm$^{-1}$, $m \alpha^2 /2  = 80 $ meV and $a=5$ nm.}
\label{BelowWavefunction}
\end{center}
\end{figure}

 All these regimes are depicted in figures \ref{BelowAsymmetry1} and \ref{BelowAsymmetry2} for the polarization as a function of the barrier length and the SO energy equivalent $m\alpha^{2}/2$. The range chosen of the SO energy is in agreement with the values computed in ref.\cite{VarelaHydrogenBond} due to hydrogen bonding. Figure~\ref{BelowAsymmetry1} shows the situation deep below the barrier where the wavefunction oscillates and decays (see Fig. \ref{BelowWavefunction}) in an under-damped situation because of the SO coupling. At zero SO coupling no spin polarization is observed. Once we have a finite $\alpha$ the polarization is exponentially enhanced but there are also interference effects due to different oscillation frequencies of the $|\uparrow,\downarrow\rangle$ spin components. This gives a reentrant effect where polarization can increase and then decrease as a function of the barrier width. Note the polarization can increase a factor of three for a change in between $0.1$ and $1$ nm in barrier length. {\color{black} At $1$ nm barrier length and 40 meV Rashba coupling (not capped by the atomic SO coupling because it is a combination of SO and Stark interactions\cite{VarelaHydrogenBond} for DNA and Oligopeptides) we find a polarization of ~30\%.} 
 
 Spin filtering by tunneling in spin active media, generates a high polarization with the expected molecular SO coupling, the amplitude is also exponentially small. Experimental accounts for the polarization rates should be able to check for this feature in time resolved experiments or essays that can change the tunneling length by e.g. mechanical stretching\cite{Chimia,VarelaHydrogenBond}.

Figure~\ref{BelowAsymmetry2} depicts a different regime where one has an input energy close to the barrier height. There we see a stronger re-entrant effect that extends for even lower values of the SO energy while increasing the needed barrier lengths for the same polarization enhancement as in Fig.\ref{BelowAsymmetry1}. The figure also shows the expected transition to plane wave behavior at $|q_0^2 - k^2|\sim(m\alpha/\hbar)^2$ under the barrier, because of the SO energy scale.  

 \begin{figure}[H]
\begin{center}
\includegraphics[width=8.5cm]{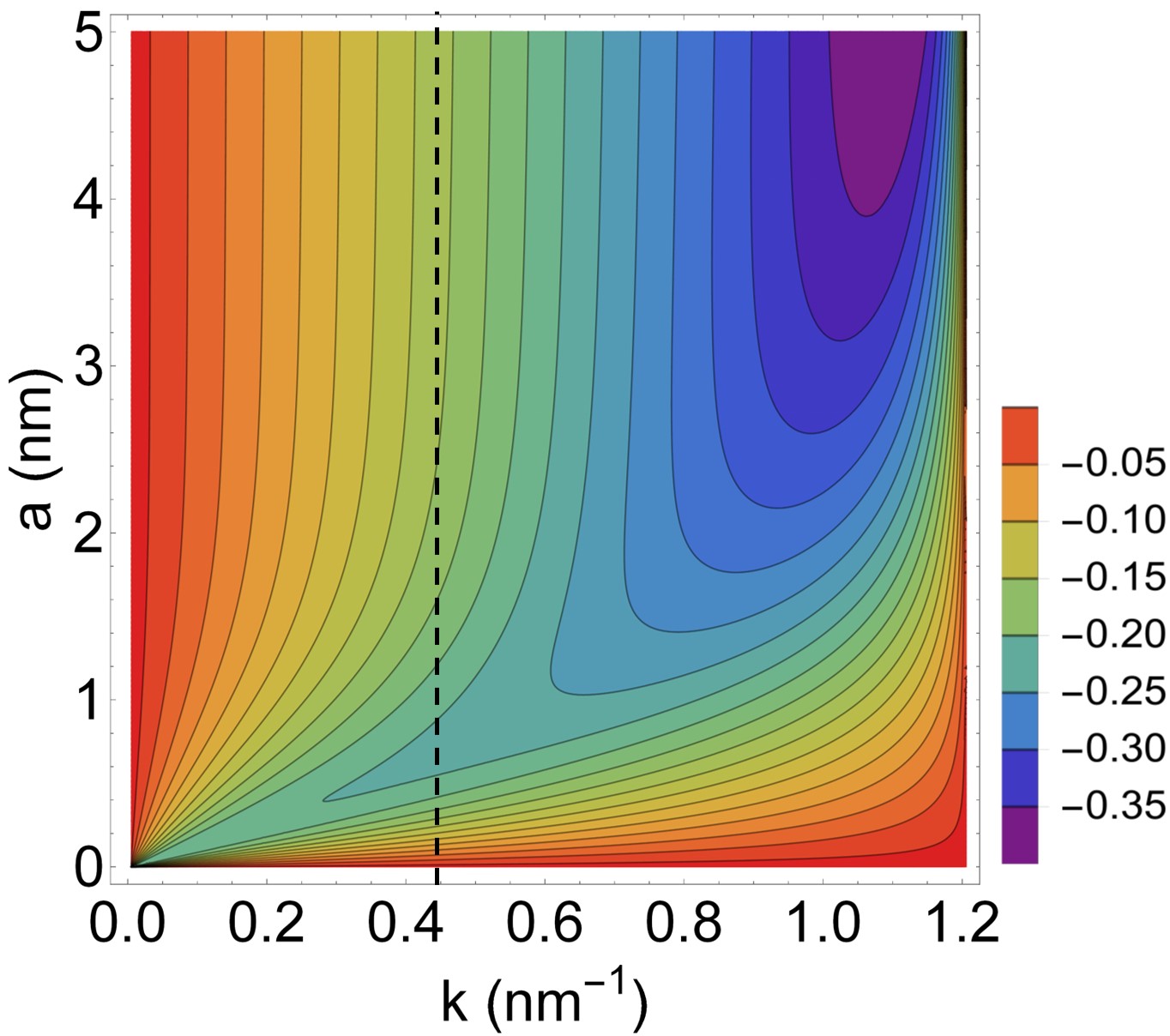}
\caption{Spin asymmetry $P_z$ versus the barrier length $a$ and the input momentum $k$, where the  barrier height $q_0=1.2$ nm$^{-1}$, and the SO coupling energy $m\alpha^2/2=30$ meV are fixed. The dotted line represents a reasonable value for $k$ argued in the text. Note the possibility of tuning between the barrier length and the input momentum determined by the pre-barrier well.}
\label{BelowAsymmetry3}
\end{center}
\end{figure}

{\color{black}Finally Fig.\ref{BelowAsymmetry3} shows the sensitivity of the barrier polarizing strength as a function of the input momentum (determined by the input well states). The figure also shows the possibility of tuning the well associated momentum and the barrier length to achieve large filtering efficiencies}. The existence of this mechanism for filtering could be evidenced by stretching/compressing the molecule in order to modify the tuning parameters and thus the filtering power of the system.

\subsection{Above barrier energies $E>V_0$}

\begin{figure}[H]
\begin{center}
\includegraphics[width=8.8cm]{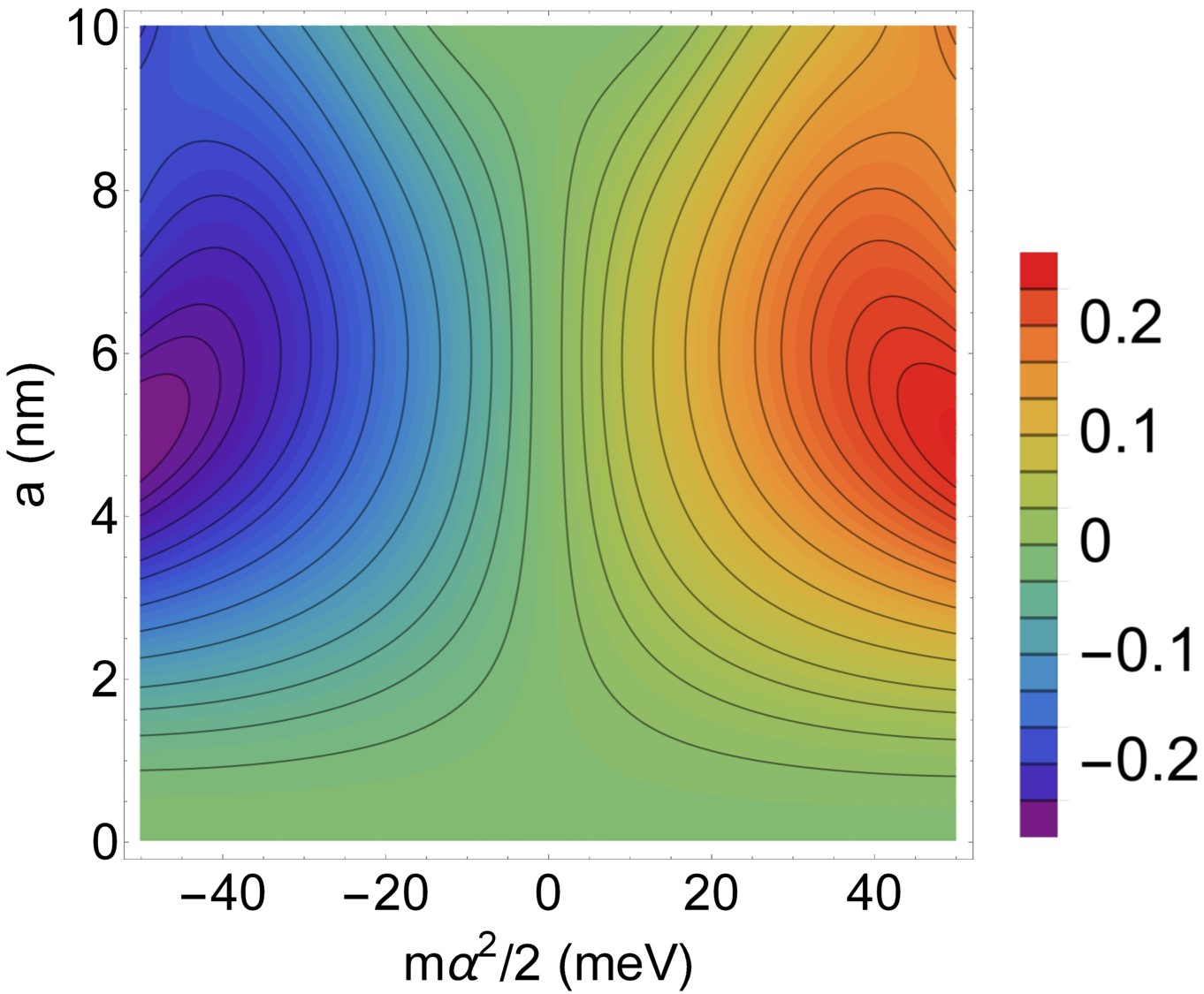}
\caption{Spin asymmetry $P_z$ as a function of $a$ in nm and $\alpha$ in meV. The values for the incident energy of electron $k=0.44$ nm$^{-1}$ and the barrier height $q_0=0.30$ nm$^{-1}$ are fixed.}
\label{Asymmetry1}
\end{center}
\end{figure}

The range of energies above the potential barrier are dominated by "interference" polarization as shown by the reentrant plot in Fig.\ref{Asymmetry1}. Here there is no exponential decay and polarization is produced by the relative oscillations of the two spin amplitudes. We believe this is not a generic situation for electron transfer in molecules where tunneling is predominant. If the energy is close to the barrier height one spin component can have energies below the barrier while the other is above the barrier and polarization can be enhanced by the same mechanism as in Fig~\ref{BelowAsymmetry2}. From the figure we can also see that the interference mechanism is less effective in producing high polarization values (up to 20\%).

\subsection{Spin currents and torque dipoles}

It has been shown that in the presence of SO coupling the conventional definition of spin current as a matrix element of $\mathbf{\hat{J}_s}=(1/i\hbar)\{\mathbf{v},{s_z}\}$ is incomplete and unphysical\cite{Shi}. The consistent spin current density should be written in the form: 
\begin{equation}
\mathfrak{I}_s=\Re{\Psi^{\dagger}(\vec{r})\mathfrak{\hat{I}}_s \Psi(\vec{r})},
\end{equation}
where $\mathfrak{\hat{I}}_s={d(\hat{r}\hat{s}_z)}/{dt}$  is the effective spin current operator, and $\Psi(\vec{r})$ is the spatially dependent wave function. Developing the definition of the conserved spin current we have
\begin{eqnarray}
 \mathfrak{\hat{I}}_s&=&\frac{d\mathbf{\hat{r}}}{dt}\hat{s}_z + \mathbf{\hat{r}}\frac{d\hat{s}_z}{dt},\nonumber\\
 &=&\frac{1}{i\hbar} \left([\mathbf{\hat{r}},\hat{H}]\hat{s}_z + \mathbf{\hat{r}}[\hat{s}_z,\hat{H}] \right),\nonumber\\
 &=& \mathbf{\hat{J}_s}+\mathbf{\hat{P}_{\tau}},
 \label{SpinPlusTorque}
\end{eqnarray}
where $\hat{H}$ is the Hamiltonian of the system, $\hat{s}_z$ is the spin operator for the $z$ component, {\color{black}$\mathbf{\hat{J}_s}$ is the conventional spin current operator and the extra term $\mathbf{\hat{P}_{\tau}}$ is the torque dipole density from the corresponding torque density $\tau$ due to the presence of the SO coupling}. 

\begin{figure}
\begin{center}
\includegraphics[width=8.8cm]{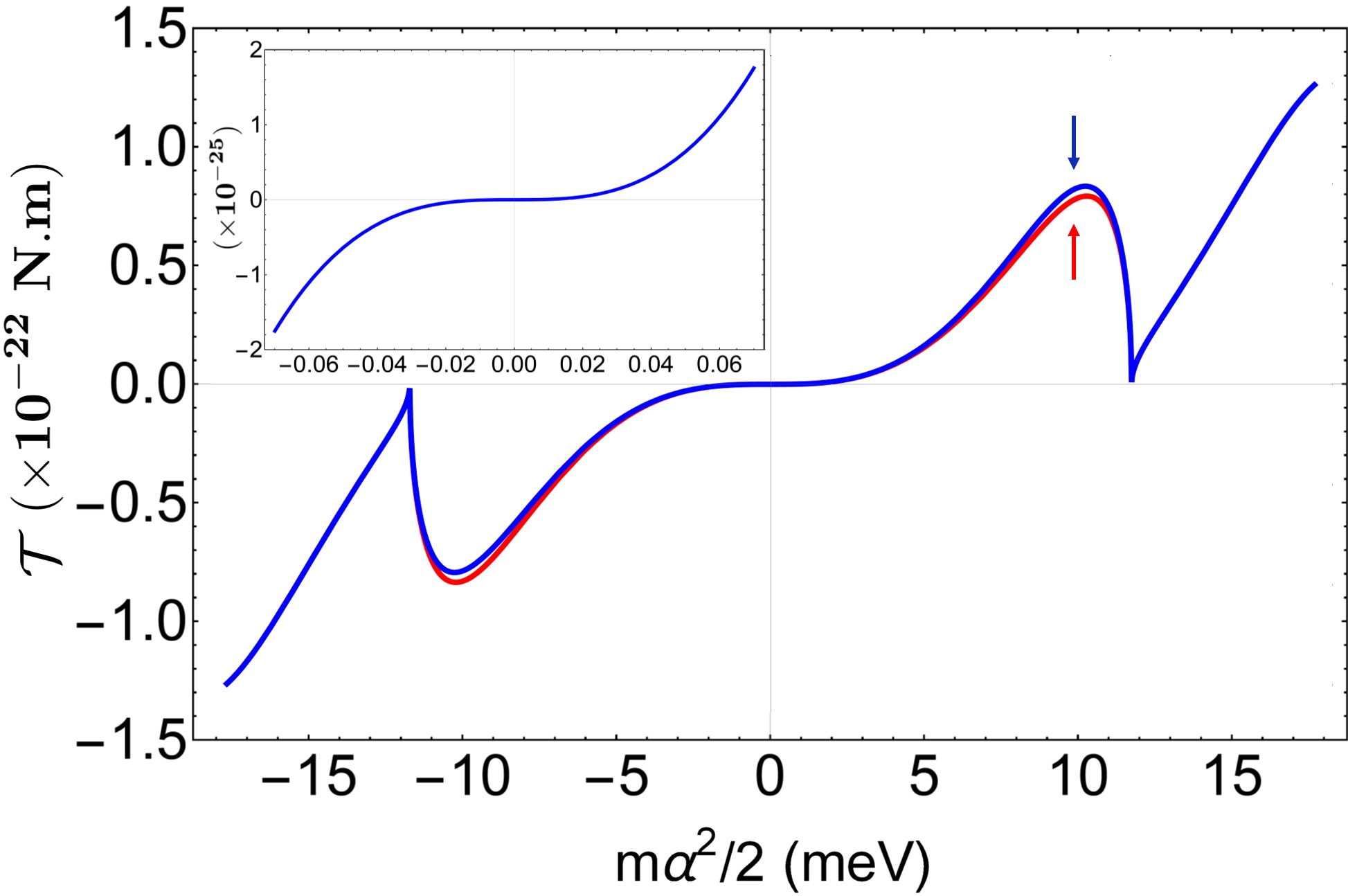}
\caption{Torque density $\tau_2$ in the region 2  for the two spin-components as a function of the SO coupling energy with $k=0.440$ nm$^{-1}$, $q_0=0.446$ nm$^{-1}$, and $a=5$ nm. Note there is no linear regime (see inset) for spin filtering. }
\label{Torque1}
\end{center}
\end{figure}

\begin{figure}
\begin{center}
\includegraphics[width=8.5cm]{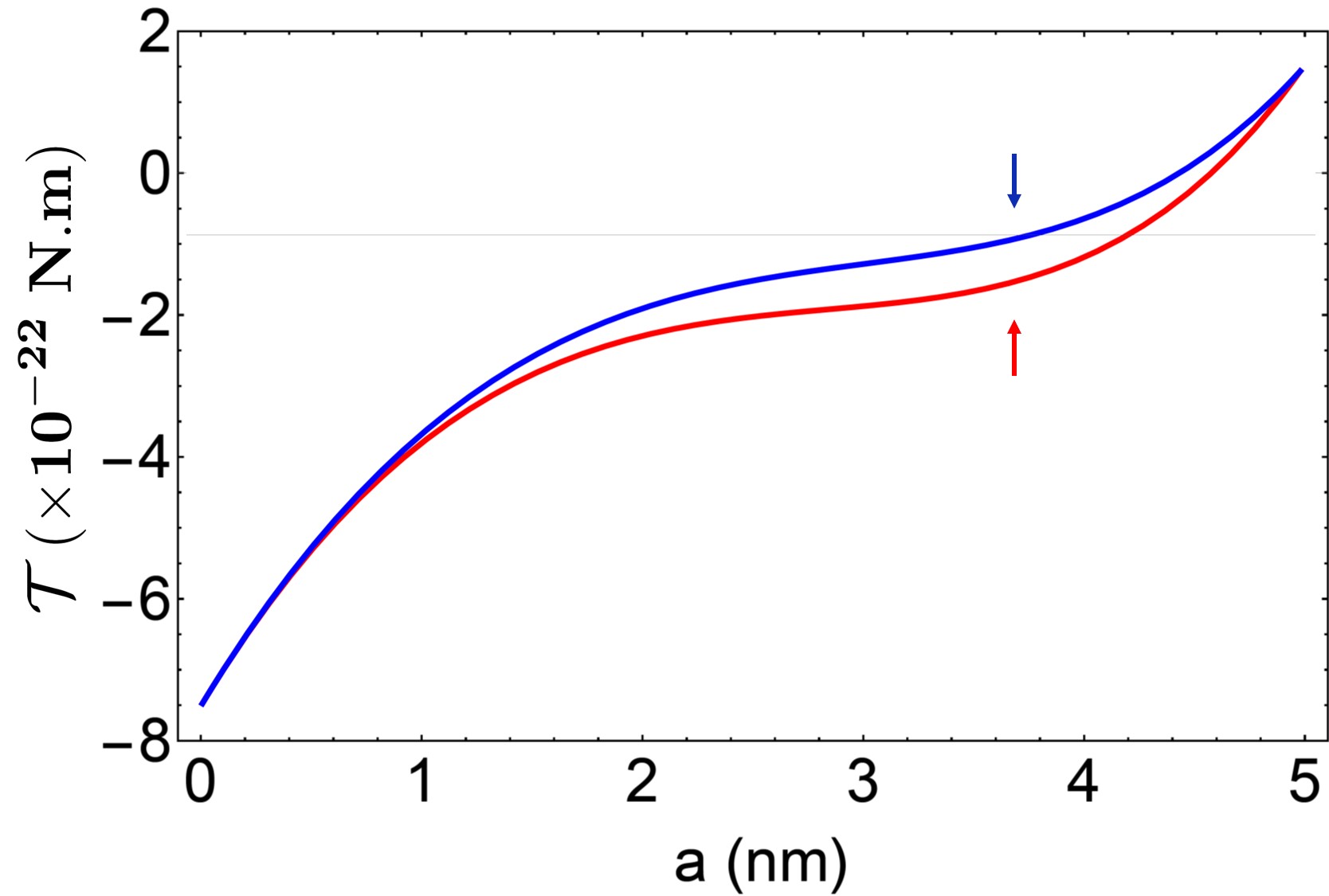}
\caption{Torque density $\mathcal{T}$ in the barrier region for the two spin-components, as a function of the width barrier with $k=0.440$ nm$^{-1}$, $q_0=0.446$ nm$^{-1}$ and $m{\alpha}^2/2=14$ meV.}
\label{Torque2}
\end{center}
\end{figure}

Considering our Hamiltonian ($\ref{HamiltonianFull}$), the two terms in Eq.\ref{SpinPlusTorque} are
\begin{equation}
\mathbf{\hat{J}_s}=\frac{-i\hbar^2}{2m}\left(\begin{array}{cc}
\partial_x & -m\alpha/\hbar\\
m\alpha/\hbar & -\partial_x
\end{array}\right),
\end{equation}
and 
\begin{equation}
\mathbf{\hat{P}_{\tau}}=ix\alpha\hbar \left(\begin{array}{cc}
0 & \partial_x\\
\partial_x & 0
\end{array}\right).
\end{equation}
The torque density can be then computed by the relation
\begin{equation}
\mathbf{\mathcal{T}_s}=\Re\left\{\Psi^{\dagger}\frac{ds_z}{dt}\Psi\right\}=\Re\left\{\Psi^{\dagger}\frac{1}{i\hbar}[{\hat s}_z,{\hat H}]\Psi\right\}=\div {\mathbf P}_s,
\end{equation}
Figure~\ref{Torque1} shows the torque density integrated over the barrier length as a function of physical values for the {\color{black}SO energy}. The figure shows the range where there is a torque differential between spin species producing net spin polarization seen previously. The sharp dip indicated the SO coupling that produces pure wave behavior under the barrier ($q_s$ purely imaginary, see Eq.\ref{wavevectorI}). It is curious to note also (see inset), there is no linear regime for small $\alpha$ that shows spin polarization. Figure~\ref{Torque2} shows similar behavior as a function of the barrier length. Again there is no linear regime for polarized currents. One can think of torques taking away angular momentum depending on the spin species as the mechanism for generating spin polarization under the barrier. This is a very clear insight derived from the consistent formulation of the conserved spin current definition\cite{Shi}.

{\color{black}As a concrete estimate of the change in angular momentum produced by the torque density: Using the input $k$ vector range in Fig.\ref{BelowAsymmetry3} to estimate the barrier dwell time \cite{Buttiker} which for $k=0.44$ nm$^{-1}$ is $~10^{-14}$ (see reference \cite{DwellTimes}). From this estimate we can compute, from Fig.\ref{Torque2}, the total change in angular momentum is $\Delta L\sim 0.1 \hbar/2$. This is a polarization that is comparable to that reported in Fig.\ref{BelowAsymmetry3}.}

\section{Summary and conclusions}
We have derived a Hamiltonian for a model of doped DNA that includes a SO coupling term that depends linearly on crystal momentum. We assume that electrons tunnel under a barrier of length $a$ between confined electron-phonon/polaron states. The SO couples differently to each component of the spinor yielding a net spin-polarized output. The output polarization can be very large, e.g. 60\% for realistic values of the SO coupling\cite{VarelaHydrogenBond}, depending on the relation between the barrier length and the input $k$ vector of the electron. This is of course at the cost of a small spin current amplitude. We have also discussed the source of spin polarization as due to the existence of a torque density that differentiates between up and down spin, using a consistent formulation of the spin current\cite{Shi}. {\color{black} This mechanism is checked with an estimate of the change in angular momentum of the electron this torque density produces}. Thus there is no need to invoke large unphysical SO strengths to achieve large polarization values, as measured in the experiments. A final feature that bears out of the model is that spin filtering has no linear regime as a function of the SO strength and the barrier length. {\color{black} These results seem to offer an alternative interpretation to models that require time reversal symmetry breaking e.g. wave function leakage to explain spin polarization in the context of the CISS effect\cite{Balseiro}}. 

{\color{black} One important conclusion related to the generality of the model is its validity for very general sequences of DNA and Oligopeptides as long as transport the mechanism involves short range tunneling\cite{Giese}. The tunneling mechanism for transport is present both in uniform and heterogeneous sequences that have been studied\cite{SequenceObservation}. Given the latter the  physically relevant ingredients in the minimal model are: a linear in $k$ SO coupling with meV strength due to C/N atoms and a consistent conserved spin-current definition providing an angular momentum changing torque density.}

\acknowledgements{This work was supported
by CEPRA VIII Grant XII-2108-06 {\it Mechanical Spectroscopy} funded by CEDIA, Ecuador. We acknowledge useful discussions with Ji\v{r}\'{i} Svozil\'ik.}


%


\end{document}